\documentclass{PoS}

\title{Challenges in Hadronic Form Factor Calculations}

\ShortTitle{Challenges in Hadronic Form Factor Calculations}

\author{\speaker{Huey-Wen Lin}\\
        Thomas Jefferson National Accelerator Facility, Newport News, VA 23606\\
        E-mail: \email{hwlin@jlab.org}}

\author{Saul D. Cohen, Robert G. Edwards, Kostas Orginos, David G. Richards\\
        Thomas Jefferson National Accelerator Facility, Newport News, VA 23606\\}

\abstract{There is an extensive history of form factor calculations on the lattice, primarily with ground states for both initial and final states. However, there have never been any radially excited transition form factor calculations. Furthermore, the lattice faces difficulty in extracting signal from noise at large transfer momenta ($Q^2$). These measurements could give important theoretical input to experiments, such as those of JLab's 12~GeV program and studies of deformation of the nucleon. In this work, I will present a simple technique to resolve both of these difficulties and present results from anisotropic configurations showing improved signals for excited-state quantities. It should also be possible to apply this technique to isotropic lattices for calculating large-$Q^2$ form factors.
}

\FullConference{The XXVI International Symposium on Lattice Field Theory\\
		 July 14-19 2008\\
		 Williamsburg, Virginia, USA}

\begin{document}

\vspace{-1cm}
\section{Introduction}
\vspace{-0.3cm}


Higher-$Q^2$ data will help us to understand hadrons and challenge QCD-based models. At Jefferson Lab, experiments are looking for whether (and if so, where) the Sachs electric form factor of the proton ($G_E^p$) crosses zero at large $Q^2$. The future 12~GeV upgrade will provide data at even larger values of $Q^2$. We in the lattice community would like to make predictions using QCD to compare with these new data sets.

The form factors of excited states are another interesting subfield being explored. Experiments at Jefferson Lab (CLAS), MIT-Bates, LEGS, Mainz, Bonn, GRAAL, and Spring-8 measured helicity amplitudes of the nucleon-Roper transition,
and many models fail to predict its value. We are particularly interested in resolving the mysterious nature of the Roper, which is still not fully understood.
Many believe that the Roper is simply the first radially excited state of the nucleon; we can verify this by providing the first-principles theoretical inputs to compare with experimental data.

Lattice QCD has successfully calculated many hadronic form factors and provided first-principles theoretical inputs for exploring the structure of hadrons. However, not much progress have been made in transition form factors or at large $Q^2$.
The typical $Q^2$ range in lattice calculations of nucleon form factors is $<3.0\mbox{ GeV}^2$; when one attempts higher-$Q^2$ calculations, they suffer from poor signal-to-noise ratios\cite{Hsu:2007ai}. In the case of transition form factors to radially excited states, the difficulty arises through the rapid exponential fall-off in Euclidean time of the excited state. By using anisotropic lattices, excited-state signals may be improved dramatically compared with conventional isotropic lattices. An exploratory study can be found in Ref.~\cite{Lin:2008qv}. In this work, we demonstrate that we can address these two lattice QCD challenges on the anisotropic lattice for the nucleon; the same techniques can be applied to other hadrons.

\vspace{-0.3cm}
\section{Methodology and Setup}
\vspace{-0.3cm}

We use quenched $16^3 \times 64$ lattices with anisotropy $\xi=3$ (that is, $a_s=3a_t$), using Wilson gauge action with $\beta=6.1$ and stout-link smeared\cite{Morningstar:2003gk} Sheikholeslami-Wohlert (SW) fermions\cite{Sheikholeslami:1985ij} with smearing parameters $\{\rho,n_\rho\}=\{0.22,2\}$. The parameter $\nu$ is nonperturbatively tuned using the meson dispersion relation, and the clover coefficients are set to their tadpole-improved values. The inverse spatial lattice spacing is about 2~GeV, as determined by the static-quark potential, and the simulated pion masses are about 480, 720 and 1100~MeV. In total, we use 200 configurations at each pion mass.

The interpolating fields we use for the nucleon have the general form for most octet baryons,
\begin{eqnarray}
 \chi^B (x) &=& \epsilon^{abc} [q_1^{a\mathrm{T}}(x)C\gamma_5q_2^b(x)]q_1^c(x),
 \label{eq:lat_B-op}
\end{eqnarray}
where $C$ is the charge conjugation matrix, and $q_1$ and $q_2$ are any of the quarks $\{u,d,s\}$. In the case of the proton, we want $q_1=u$ and $q_2=d$.

\begin{figure}
\begin{center}
\includegraphics[width=0.45\textwidth]{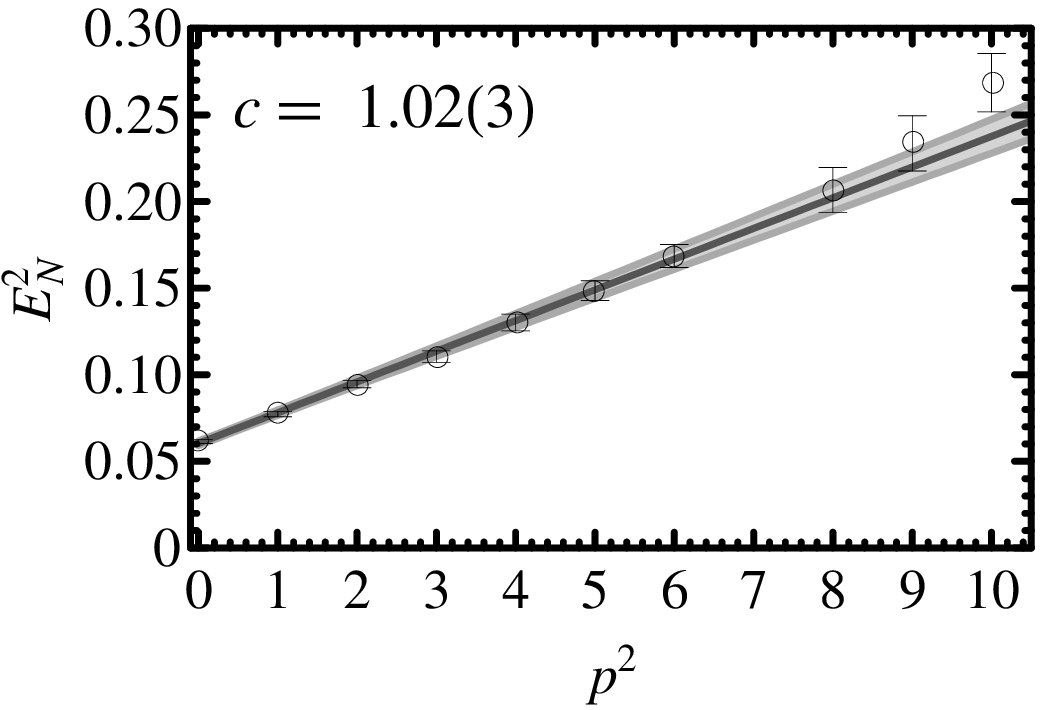}
\includegraphics[width=0.45\textwidth]{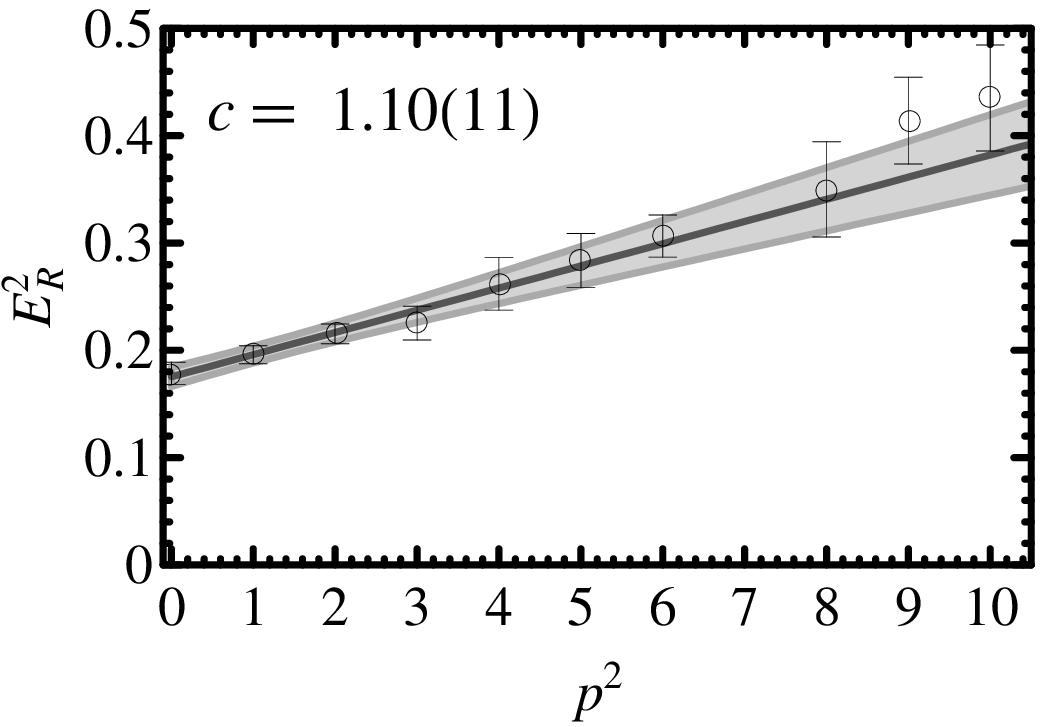}
\end{center}
\vspace{-0.7cm}
\caption{Nucleon and $P_{11}$ dispersions. The y-axis is in units of $a_t^{-2}$, while the x-axis is in units of $\frac{4\pi^2}{L_{x,y,z}^2}a_s^{-2}$. Both figures show slopes consistent with 1.}\label{fig:NR-disp}
\end{figure}

At each pion mass, we use three Gaussian smearing parameters: $\sigma \in \{0.5, 2.5, 4.5\}$; the largest of these has excellent overlap with just the ground state, while the other two include substantial higher-state contributions. For both two-point and three-point correlators, we calculate all 9 possible source-sink smearing combinations. We apply the variational method\cite{Luscher:1990ck} to extract the principal correlators corresponding to pure energy eigenstates from our matrix of correlators. This gives 3 principal correlators; we discard the $2^{\rm nd}$-excited correlator which is likely contaminated by unresolved higher states. We extract the energies of the ground and first-excited nucleon by applying simple exponential fits individually to the lowest two principal correlators. After this, we also check whether the original two-point correlators are recovered by appropriate combinations of the exponentials. Figure~\ref{fig:NR-disp} shows the the dispersions of the nucleon and its first radially excited state at $m_\pi\approx720$~MeV; they both have slopes consistent with the expected continuum value, 1. Similar results also found for the other two pion masses. The excited state has a better signal compared with those obtained from conventional isotropic lattices.

To calculate three-point functions, we choose source and sink locations to be at timeslices 15 and 48 respectively. (This gives a source-sink separation about 1.1~fm.) We choose the final state to always have zero momentum in this proceeding, while the momentum of the initial state varies.

The three-point function has the form
\vspace{-0.1cm}
\begin{eqnarray}\label{eq:general-3pt}
\Gamma^{(3),T}_{\mu,AB}(t_i,t,t_f,\vec{p}_i,\vec{p}_f) &=&
Z_V \sum_n \sum_{n^\prime} f_{n,n^\prime}(p_f,p_i,E_n^\prime,E_n,t,t_i,t_f)\nonumber \\
&\times& \sum_{s,s^\prime}
T_{\alpha\beta} u_{n^\prime}(\vec{p}_f,s^\prime)_\beta
\langle N_{n^\prime}(\vec{p}_f,s^\prime)\left|V_\mu\right|N_n(\vec{p}_i,s)\rangle\overline{u}_n(\vec{p}_i,s)_\alpha,
\end{eqnarray}
\vspace{-0.1cm}
where $f_{n,n^\prime}(p_f,p_i,E_n^\prime,E_n,t,t_i,t_f)$ contains kinematic and overlap factors and time-dependent terms, $n$ and $n^\prime$ index different energy states, the projection used is $T=\frac{1}{4}(1+\gamma_4)(1+i\gamma_5\gamma_3)$, and $Z_V$ is the vector-current renormalization constant (which is set to its nonperturbative value).
The vector current in Eq.~2.2 is $O(a)$ on-shell improved with the improved coefficient set to its tree-level value. Note that when calculating the three-point Green function in full QCD, there are two possible contraction topologies: ``connected'' and ``disconnected'', when the vector current vertex appears on a vacuum bubble. In this work, only ``connected'' quantities are included.

The ratio method will not extract matrix elements beyond the ground state, but using the $Z$'s and $E$'s derived from our analysis of the two-point functions, we can extract excited matrix elements from the three-point functions by fitting to the form given in Eq.~2.2. We want to keep terms in Eq.~2.2 having $n$ and $n^\prime$ running from the ground to the first-excited state; thus, there would be four matrix elements in the minimal expansion, which will be free parameters in our fit. We increase the number of three-point correlators, first using just the diagonal correlators where the smearing is the same at the source and sink, and then using all 9 smearing combinations. The nucleon-nucleon matrix elements are verified against the ratio method, and we check that the transition matrix elements are consistent between different sets. In the rest of this work, we show the results from full 9-correlator simultaneous fits.

The vector-current matrix elements, $\langle N\left|V_\mu\right|N\rangle$ (with $n=n^\prime=0$) and $\langle N\left|V_\mu\right|P_{11}\rangle$ (with $n=0$, $n^\prime=1$) (or its conjugate) are fitted according to Eq.~2.2. The fit ranges are carefully chosen to give consistent results. The form factors and the vector-current matrix element for any nucleon states, $N_1$ and $N_2$, are connected through
\begin{eqnarray}
\label{eq:Vector-roper}
\langle N_2\left|V^{\rm }_\mu\right|N_1\rangle_{\mu}(q) &=&
{\overline u}_{N_2}(p^\prime)\left[ F_1(q^2) \left(\gamma_{\mu}-\frac{q_\mu}{q^2}
q\!\!\!\!/
\right)
\sigma_{\mu \nu}q_{\nu}
\frac{F_2(q^2)}{M_{N_1}+M_{N_2}}\right]u_{N_1}(p) e^{-iq\cdot x},
\end{eqnarray}
where the equation of motion is used to simplify $-q_\mu \gamma_\mu=M_{N_2}-M_{N_1}$. By using singular-value decomposition (SVD), we solve for the form factors $F_{1,2}$.

\vspace{-0.3cm}
\section{Numerical Results}
\vspace{-0.3cm}

\begin{figure}
\begin{center}
\includegraphics[width=0.5\textwidth]{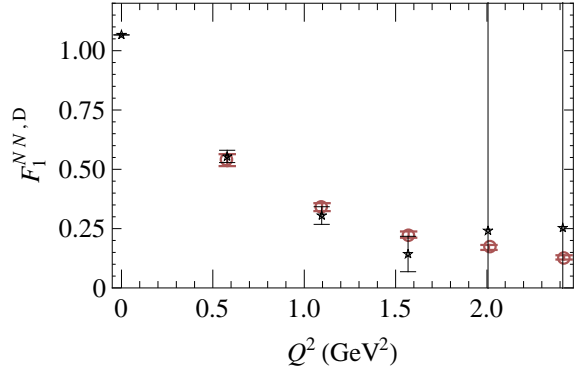}
\end{center}
\vspace{-0.7cm}
\caption{$d$ quark contribution to the nucleon Dirac form factor $F_1^d$, obtained from the ratio approach (stars) and the fitting method (circles)}\label{fig:fNN-comp}
\end{figure}

The ground-state nucleon form factors, $F_{1}^{NN}$ and $F_{2}^{NN}$, have been well studied, and we can use the conventional analysis method to crosscheck our fitting proposal. We use the largest Gaussian smearing ($\sigma=4.5$) three-point data, which has good overlap with ground states; this makes it ideal for use in a ratio approach\cite{Hagler:2003jd} to get nucleon form factors. We obtain nucleon-nucleon form factors $F_1$ and $F_2$ consistent with those derived from the ratio method. Fig.~\ref{fig:fNN-comp} shows that both methods have consistent $d$-quark contributions to the Dirac form factor at pion mass 720 MeV. Note that we not only get consistent numbers in $F_1$, but at large momenta, our fitting approach dramatically improves the signal. This is because when one tries to improve the ground-state signal (normally done by examining the hadron effective-mass time dependence at zero momentum) with a choice of smearing operator, one wipes out not only the excited states but also the higher-momentum states. Therefore, when one tries to project onto these higher momenta, there is no doubt that the signal-to-noise ratio will worsen. Since we are explicitly considering excited states in our analysis, our sources which have good overlap with higher-momenta will be only lightly contaminated by excited-state signals.

\begin{figure}
\includegraphics[width=0.5\textwidth]{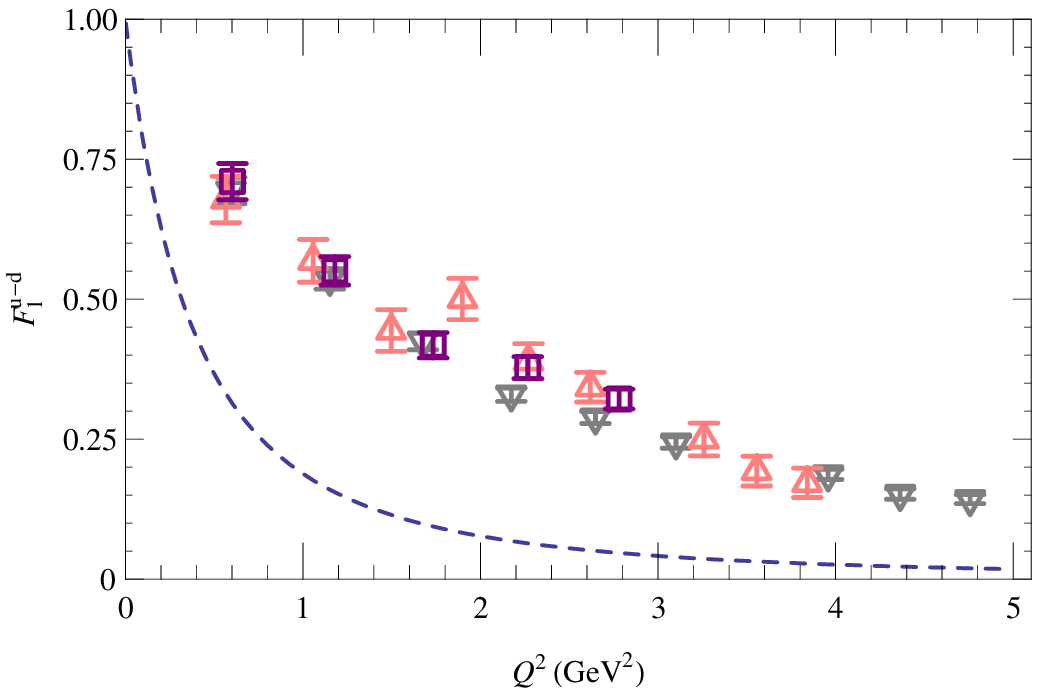}
\includegraphics[width=0.5\textwidth]{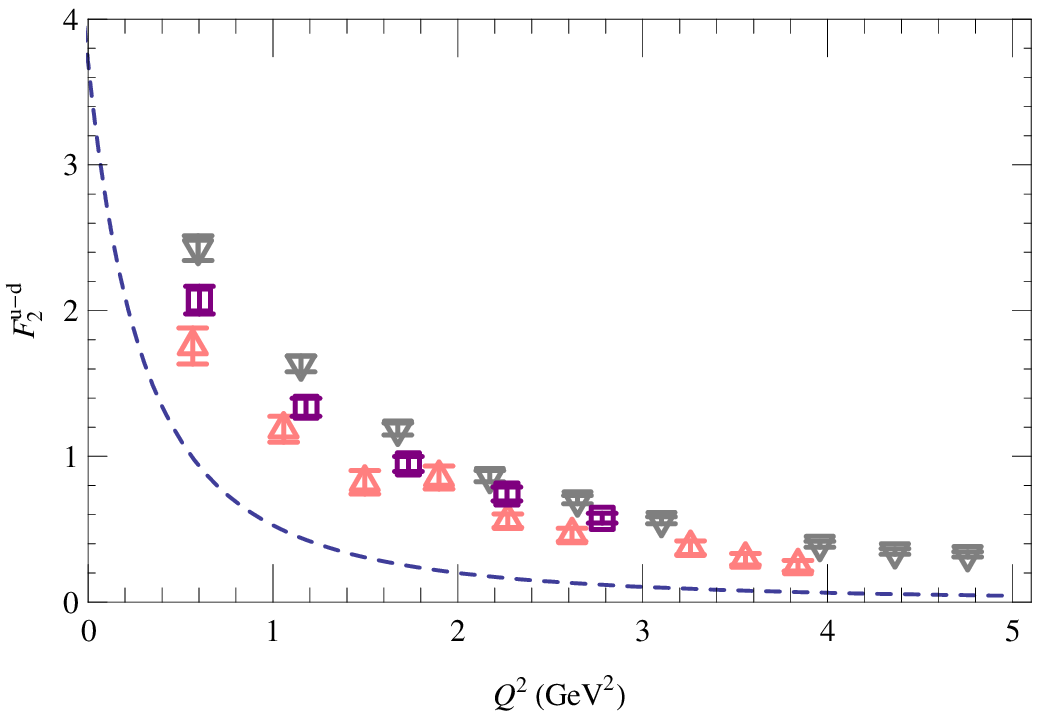}
\vspace{-0.7cm}
\caption{Isovector Dirac (left) and Pauli (right) nucleon form factors, $F_1$ and $F_2$, with pion masses of 480 (pink), 720 (purple) and 1100 (grey) MeV. The dashed lines indicate the experimental values.
}\label{fig:isoFF}
\end{figure}

Figure~\ref{fig:isoFF} shows preliminary results with isovector Dirac and Pauli form factors, $F_{1,2}$ with three values of light-quark mass, which correspond to roughly 480 (pink triangles), 720 (purple squares) and 1100 (downward triangles) MeV pions. We see very mild pion-mass dependence over the entire $Q^2$ range for the Dirac for factor and in the large-$Q^2$ region for the Pauli form factor. Although $F_2$ shows a small tendency toward the experimental curve, these form factors still deviate significantly from experiment. This is somewhat expected, since our pion masses are much heavier than the physical value. Even in unquenched calculations, the nucleon form factors do not agree with experimental values with pion masses as low as 300~MeV. (See reviews in Ref.~\cite{Perdrisat:2006hj} or these proceedings\cite{Zanotti:2008}.) We may expect to see the lattice data approaching experiment as the pion mass used is decreased in future calculations. Further study below $300$~MeV sea pion mass will be critical to understanding this behavior.

Note that we get very clean signal in the range $3\mbox{ GeV}^2 < Q^2 < 5$~GeV$^2$. When we examine the electric Sachs form factor
\begin{eqnarray}
G_E (q^2) &=& F_1(q^2) - \frac{q^2}{4M_N^2}F_2(q^2),
\end{eqnarray}
we see that $G_E^p$ does not cross zero within our $Q^2$ region. In the near future, we will continue analyzing data with non-zero final-state momentum to look for even larger $Q^2$ form factors.

\begin{figure}
\includegraphics[width=0.5\textwidth]{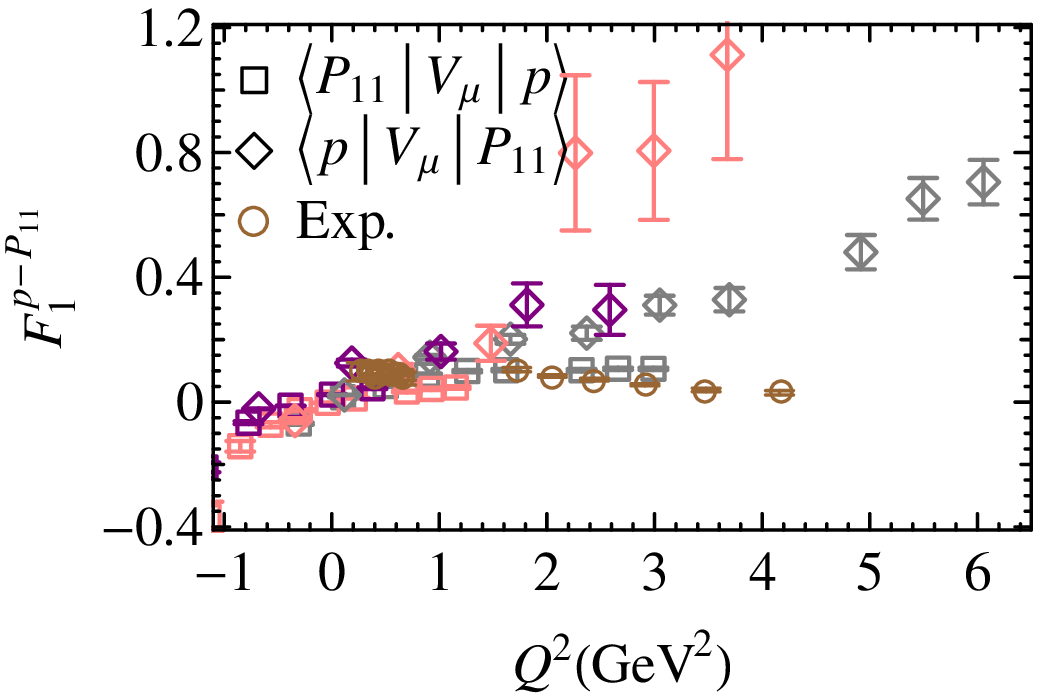}
\includegraphics[width=0.5\textwidth]{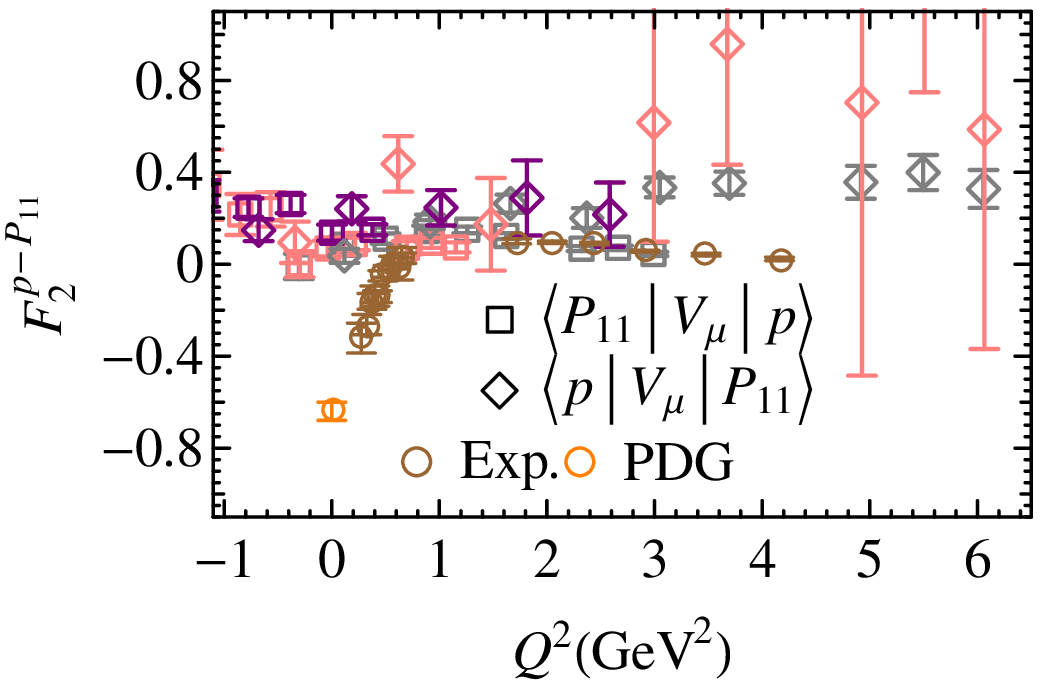}
\vspace{-0.7cm}
\caption{Proton-$P_{11}$ transition form factors: $F_1^{p-P_{11}}$ (left) and $F_2^{p-P_{11}}$ (right) with pion masses of 480 (pink), 720 (purple) and 1100 (grey) MeV. The brown points are the experimental results from CLAS and the orange are numbers from the PDG. Two matrix elements are used to extract the form factors: $\langle p\left|V_\mu\right|P_{11}\rangle$ (diamonds) and $\langle P_{11}\left|V_\mu\right|p\rangle$ (squares).}\label{fig:F12-pR}
\end{figure}

\begin{figure}
\includegraphics[width=0.5\textwidth]{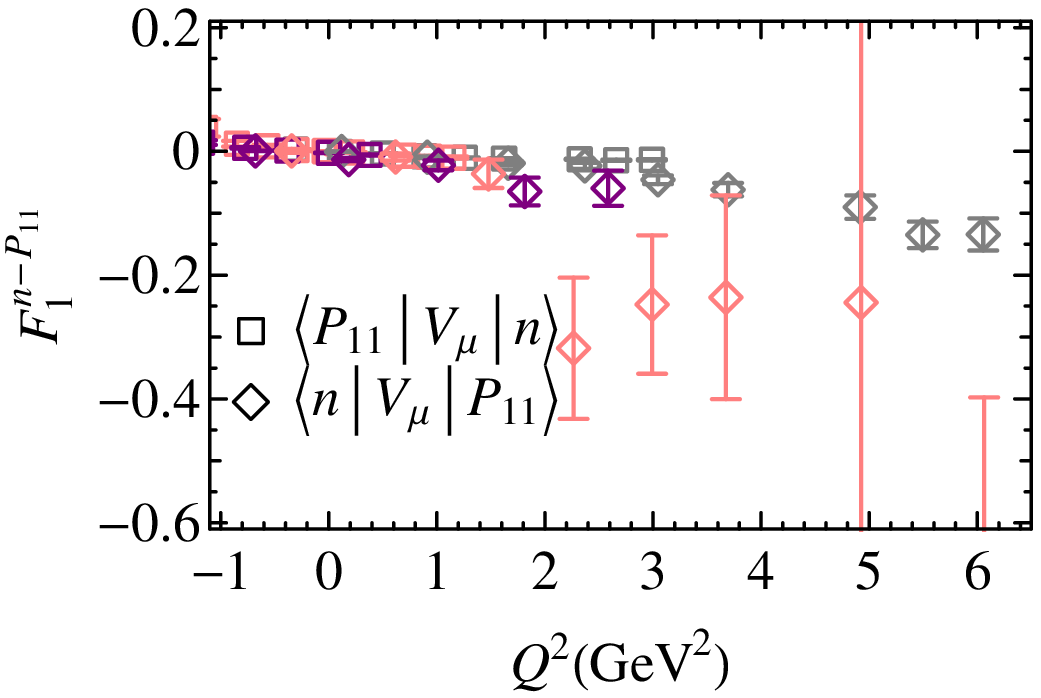}
\includegraphics[width=0.5\textwidth]{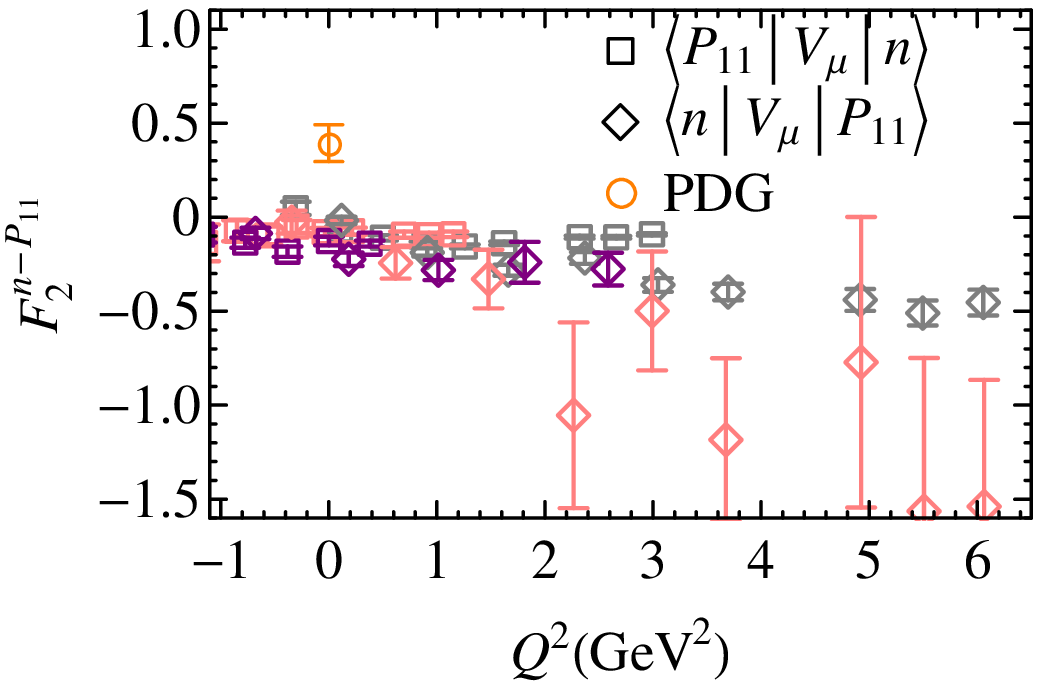}
\vspace{-0.7cm}
\caption{Neutron-$P_{11}$ transition form factors: $F_1^{n-P_{11}}$ (left) and $F_2^{n-P_{11}}$ (right) with pion masses of 480 (pink), 720 (purple) and 1100 (grey) MeV.
The symbols are the same as in Figure~4. 
}\label{fig:F12-nR}
\end{figure}

We can obtain the $P_{11}$-$N$ transition form factors from processes that are related to the Roper decay $P_{11} \rightarrow \gamma N$ ( $\langle p\left|V_\mu\right|P_{11}\rangle$) and the one that is related to photoproduction, $\gamma^* N \rightarrow P_{11}$ ( $\langle P_{11}\left|V_\mu\right|p\rangle$). We show preliminary results for transition form factors $F_{1,2}^*$ at three pion masses in Figure~\ref{fig:F12-pR} and Figure~\ref{fig:F12-nR}; the experimental values are reconstructed from experimental helicity amplitudes using single-pion or two-pion channel analyses\cite{Aznauryan:2008pe,Aznauryan:2007pi,Park:2007tn,Mokeev:2006an}
and PDG\cite{PDBook}.
Note that since our nucleon and Roper masses are much higher than the physical ones, we are in the time-like region when we use the matrix element $\langle P_{11} |V_\mu | N \rangle$ to construct the transition form factors. As we decrease the pion mass, we will enter the space-like region, and this matrix element will be helpful in giving us different $Q^2$.

A $P_{11}$ at rest on the lattice has a decay into a pion and a nucleon in a $P$-wave and a decay into two pions and a nucleon in an $S$-wave via $S_{11}\pi$. Since we do not wish to consider the complicated case that occurs when multi-particle states may be present, we must avoid kinematical situations in which decays can occur. In our simulation, these are suppressed by the high quark mass and discretization of momentum. However, the lowest-momentum $P_{11}$ can decay into a lowest-momentum pion and a nucleon at rest. Since we cannot untangle the $P_{11}$ from the two-particle state here, the leftmost points from  $\langle p\left|V_\mu\right|P_{11}\rangle$ elements should be disregarded.

In the time-like $Q^2$ regions, all four transition form factors have very mild quark-mass dependence. Even in the space-like regions, the dependence only become evident at $Q^2>2$~GeV$^2$. We obtain form factors with momentum transfer up to 6~GeV$^2$; unfortunately, our lightest pion-mass calculation suffers from large statistical error; we need more statistics to give a clean signal at large $Q^2$.  In the large-$Q^2$ region, our calculations show that these form factors have about the same magnitude with the single-pion or two-pion analyses of the experimental data. However, in the low-$Q^2$ region, both the CLAS analysis and PDG value (at zero momentum transfer) have opposite sign from ours for both proton and neutron transition form factors. Pion cloud effects could be significantly larger in the low $Q^2$ region; therefore, removal of the quenched approximation would be very helpful for distinguishing whether there is a true discrepancy in the sign.

\vspace{-0.3cm}
\section{Conclusion}
\vspace{-0.3cm}

In this work, we try to solve two form-factor challenges in lattice QCD calculations: large-$Q^2$ form factors and excited-ground state transition form factors. We demonstrate the method using a baryonic system, which typically has worse signal-to-noise ratio than mesons. The method can easily be applied to other hadronic systems.

Using multiple operators to determine the ground and excited states, we are able to extract good signals for both the ground-state form factors at large $Q^2$ and the excited-ground transition form factors. In this work, we construct a basis from operations with various smearing parameters. We see significantly improved signal-to-noise ratios for large-$Q^2$ momentum $N$-$N$ form factors compared with previous studies. We find no zero-crossing for $G_E^p$ up to 5.5~GeV$^2$.
We calculate $N-P_{11}$ transition form factors up to around 6~GeV$^2$ and find small quark-mass dependence. We would like to move on to dynamical calculations to re-examine the low-$Q^2$ behavior. Future generalizations to operators constructed in irreducible representations of the cubic group will probably allow us to analyze form factors from more than one radially excited state.

\vspace{-0.3cm}
\section*{Acknowledgements}
\vspace{-0.3cm}
This work was done using the Chroma software suite\cite{Edwards:2004sx}, part of the propagator calculation used the EigCG solver\cite{Stathopoulos:2007zi} (implemented by Andreas Stathopoulos and Kostas Orginos for the Chroma library), and calculations were performed on clusters at Jefferson Laboratory using time awarded under the SciDAC Initiative. Authored by Jefferson Science Associates, LLC under U.S. DOE Contract No. DE-AC05-06OR23177. The U.S. Government retains a non-exclusive, paid-up, irrevocable, world-wide license to publish or reproduce this manuscript for U.S. Government purposes.

\vspace{-0.4cm}

\end{document}